\documentclass[sigconf,natbib=true]{acmart}

\AtBeginDocument{%
  }

\setcopyright{acmlicensed}
\copyrightyear{2018}
\acmYear{2018}
\acmDOI{XXXXXXX.XXXXXXX}
\acmConference[Conference acronym 'XX]{Make sure to enter the correct
  conference title from your rights confirmation email}{June 03--05,
  2018}{Woodstock, NY}
\acmISBN{978-1-4503-XXXX-X/2018/06}

\usepackage{layouts}
\usepackage{colortbl}
\usepackage{algorithm}
\usepackage{algorithmic}
\begin{document}

\title{Generative Long-term User Interest Modeling for Click-Through Rate Prediction}

\author{Jiangli Shao, Kaifu Zheng, Hao Fang, Huimu Ye, Zhiwei Liu, Bo Zhang, Shu Han, Xingxing Wang}
\affiliation{%
	\institution{MeiTuan}
	\city{Beijing}
	\country{China}}

\renewcommand{\shortauthors}{et al.}

\begin{abstract}
Modeling long-term user interests with massive historical user behaviors enhances click-through rate (CTR) prediction performance in advertising and recommendation systems. Typically, a two-stage framework is widely adopted, where a general search unit (GSU) first retrieves top-$k$ relevant behaviors towards the target item, and an exact search unit (ESU) generates interest features via tailored attention. However, current target-centered GSU would ignore other latent user interests, leading to incomplete and biased interest features. Additionally, the matching-based retrieval process in GSUs depends on the pairwise similarity score between target item and each historical behavior, which not only becomes time-consuming for online services as user behaviors continue to grow, but also overlooks the interaction information among user behaviors. To combat these problems, we propose a \textbf{Gen}erative \textbf{L}ong-term user \textbf{I}nterest model named GenLI for CTR prediction. GenLI consists of an interest generation module (IGM), a behavior retrieval module (BRM), and an interest fusion module (IFM). The IGM generates multiple interest distributions to indicate different aspects of real-time user interests, which is target-independent and incorporates interaction information among behaviors, ensuring complete and diverse interest features. The BRM selects related behaviors via a simple lookup operation, reducing the time complexity for weighting each behavior to $O(1)$. Finally, the IFM uses delicate gating mechanisms to generate interest features. Based on the generation process, GenLI improves the diversity of user interests and avoids complex matching-based behavioral retrieval, achieving a better balance between accuracy and efficiency for CTR prediction. Extensive offline and online experiments demonstrate the superior performance of GenLI. Notably, GenLI has been deployed in a real-world platform, serving the main traffic of hundreds of millions of users, showing its application value. 
\end{abstract}



\received{xx}
\received[revised]{xx}
\received[accepted]{xx}

\maketitle

\section{Introduction}
Click-through rate (CTR) prediction is a core task for online commercial platforms~\cite{at4ctr,ctr_review_ad}. In recommendation systems, accurate CTR helps deliver the most appropriate content to users, enhancing user satisfaction while increasing user engagement~\cite{wide_deep,dcnv2}. In advertising systems, CTR influences the calculation of Effective Cost Per Mille (eCPM) in the cost-per-click (CPC) billing model, thus directly affecting ad revenue~\cite{deep_ctr,ctr_adreview}. Hence, it is crucial for platforms to build an accurate CTR prediction model. The core of CTR modeling is personalization~\cite{dgin,dmin,atrank}. To accomplish this, CTR models should be highly individualized, leveraging user behavioral data to its full potential. These historical user behaviors contain user interest patterns~\cite{behav_1,behav_2}, usually serving as valid features for CTR prediction. Therefore, user interest modeling has become a hot topic in both industry and academia.

Generally, user interest modeling includes capturing short-term and long-term user interests from user behavior sequences. For short-term user interest modeling, researchers aggregate users' short-term behaviors with fine-grained attention mechanisms, such as DIN~\cite{din}, DIEN~\cite{dien}, DSIN~\cite{dsin}, etc. In long-term user interest modeling, a cascaded two-stage framework is widely adopted to balance the accuracy and inference latency~\cite{sim,ubr4ctr}. In this framework, a GSU first retrieves top-$k$ relevant behaviors towards the target item with simple similarity metrics such as BM25~\cite{ubr4ctr}, inner product~\cite{sim}, or hash fingerprint~\cite{sdim,eta}. Subsequently, based on more complicated attention mechanisms, an ESU aggregates retrieved behaviors to generate long-term user interest features. Compared with short-term interests, long-term interests provide richer information on user interest patterns and display a more complete view of user preference, leading to better CTR prediction performance in systems. Meanwhile, with the development of platforms, more user behavior data are gradually accumulated, supporting the modeling of interests over a longer range. Thus, more efforts have been made to model long-term user interests in recent years.

However, current long-term interest modeling approaches have some deficiencies that may impair performance. First, existing GSUs are target-centered, considering only target-related interests while ignoring other latent user interests. Users generally have multiple interests~\cite{cim}, and focus on just one aspect can introduce bias in interest modeling. In addition, the retrieval process in GSUs depends on the similarity score between historical behaviors and target items. This score is calculated using a matching-based paradigm, requiring comparing the target item with each historical behavior. This time-consuming process will become a bottleneck when user behaviors extend, limiting the length of user behavior sequences that can be considered in online services. Moreover, GSUs adopt a pairwise architecture that compares the target item with every single behavior, ignoring the interaction information regarding user behaviors. This results in inaccurate similarity metrics for GSUs, further degrading the CTR prediction performance.

To address these issues, we propose a generative long-term user interest modeling framework named GenLI for CTR prediction. Unlike existing methods that rely on target-centered retrieval, GenLI uses a generative approach to directly produce diverse and comprehensive user interest distributions and selects behaviors via distribution lookup to improve efficiency. In detail, GenLI is composed of an interest generation module (IGM), a behavior retrieval module (BRM), and an interest fusion module (IFM). The IGM takes the recent few behaviors as input and outputs the implicit, explicit, and relative interest distributions. These distributions collectively encapsulate the real-time user interests. Importantly, this module operates independently of any target item, ensuring that the generated distributions encompass a broader range of interests beyond just the target items. Based on each interest distribution, the BRM selects a subset of behaviors that best fit the corresponding type of interest through lookup scores. The scores describe the importance of behaviors within multiple interest distributions, reflecting diversified interests. In addition, the score incorporates interaction information within behaviors, which is more precise than simple pairwise metrics. Otherwise, the lookup process is more efficient as it avoids the comparison between the target item and each historical behavior, reducing the time complexity to weight each behavior to $O(1)$. Finally, the IFM separately aggregates the selected behaviors from different interest distributions and eventually uses gating mechanisms to produce long-term user interest features, serving as the main features for CTR prediction. Relying on the generative method, GenLI improves the diversity of user interests and avoids complex matching-based behavioral retrieval, achieving a better balance between accuracy and efficiency.

GenLI demonstrates superior performance in offline experiments and achieves a promotion of 1.56\% in RPM (Revenue per Mille) in online A/B testing. As its application value, GenLI has been deployed in online advertising systems in the real-world platform and serves hundreds of millions of customers every day. The main contributions of our work are summarized as follows:

\begin{itemize}
\item To the best of our knowledge, we are the first to introduce generative methods in modeling long-term user interests for CTR prediction tasks.

\item We propose a framework named GenLI for long-term user interest modeling, which consists of the IGM, BRM, and IFM. The IGM first generates user interest distributions. The BRM then selects behaviors that users are most likely to be interested in, and the IFM aggregates them to produce long-term interest features. GenLI is an end-to-end framework and can be easily plugged into any other CTR prediction model.

\item GenLI shows superior performance in both offline and online experiments. It has been deployed in real-world advertising systems and serves millions of users every day.
\end{itemize}

\section{Related Works}
\subsection{User Behavior Modeling}
User behavior modeling~\cite{behav_1,behav_2} has received significant attention because it extracts valuable user interest features from historical user behavior sequences for CTR prediction tasks. In early methods, hand-crafted features of user behavior sequences are precomputed offline due to limited online resources~\cite{wide_deep,deepfm,dcnv2}. Later, multiple attention mechanisms are designed to model the interaction between the target item and historical user behaviors, such as DIN~\cite{din}, DIEN~\cite{dien}, and DSIN~\cite{dsin}. Despite significant improvement, these methods are typically applied when user behaviors are less than 100 due to the high computational cost.

With the accumulation of user data, behavior sequences become longer and longer, bringing more interest patterns and more personalized user features. Thus, how to effectively capture long-term user interest features has become a hot topic. To balance the behavior sequence length and computational resource consumption, MIMN~\cite{mimn} proposes a memory-based architecture for decoupling the most resource-consuming part. Later, SIM~\cite{sim} designs a two-stage cascaded framework, which first retrieves a subset of candidates from thousands of historical user behaviors with a simple general search unit and then uses a heavy but accurate exact search unit to extract the final interest representation. In SIM, hard and soft strategies are proposed in GSU, which select historical items with the same category and the highest inner product score, respectively. Later, more GSU modules are proposed for retrieving relevant behaviors quickly and precisely. For example, UBR4CTR~\cite{ubr4ctr} filters candidate behaviors according to BM25~\cite{bm25} metrics. ETA~\cite{eta} uses locality-sensitive hash (LSH)~\cite{simhash} to encode and select behaviors. SDIM~\cite{sdim} samples behaviors with the same hash signature as the target item via multi-round hash collision. Otherwise, TWIN~\cite{twin} ensures consistency between the two stages by utilizing identical optimized multi-head attention mechanisms in both GSU and ESU, achieving SOTA performance. However, current user interest modeling methods are target-centered, mainly capturing the target-related interest but ignoring the valuable user latent interest. Thus, the captured interest is incomplete, degrading the performance.

In this paper, we also focus on how to capture long-term user interest from historical user behaviors. Different from existing works, we drew inspiration from generative methods' ability to create novel and creative results and utilized their modeling ability for historical sequences to generate real-time user interest distributions in a target-independent manner for the first time in CTR prediction tasks. The generated interest distribution can reflect complete and diverse user interests, helping to improve performance.

\begin{figure*}[!ht]
	\centering
	\includegraphics[width=\linewidth]{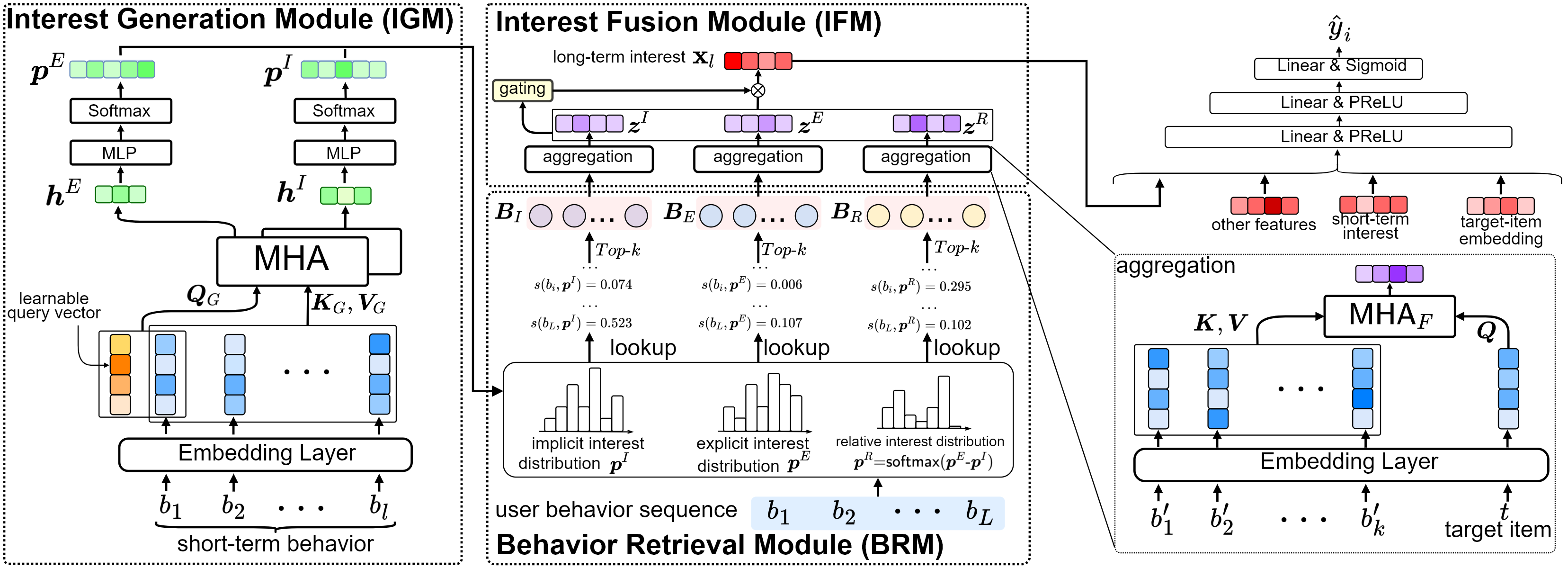}
	\caption{The framework of GenLI. GenLI consists of an interest generation module (IGM), a behavior retrieval module (BRM), and an interest fusion module (IFM).}
	\label{pic_1}
\end{figure*}

\subsection{Generative Sequence Modeling}
Generative methods have achieved considerable success in sequence modeling in various domains~\cite{gen_seq_1,gen_rec_ret}, showing their excellent ability. For instance, generative models have been extensively used in natural language processing to learn knowledge from word sequences~\cite{gpt,gen_nlp_1}. Generative information retrieval models can generate credible responses for search queries based on given document sequences~\cite{gen_ret_1,gen_ret_2,gen_ret_3}. For recommender systems, generative methods have proven to be powerful tools for capturing latent user preferences from sequences of user interactions~\cite{gpt4rec,gen_rec1}, such as purchase or browse behaviors. Also, generative methods can be utilized to recall and ranking processes in recommendations~\cite{gen_rec_ret,gen_rec2}. Generative models can uncover hidden patterns and preferences, improve the accuracy and personalization of recommendations, and lead to better user satisfaction.

However, the potential of generative methods has not yet been fully realized in CTR prediction tasks. Most existing recommendation methods focus on generating specific item IDs, which have problems with sparsity and low efficiency. In this paper, we aim to investigate using generative approaches to extract long-term user interests to enhance CTR prediction. Rather than directly generating item IDs, we aim to capture the rich and valuable interest information in historical user behavior data using a generation process, thereby developing a fast yet powerful CTR prediction model.

\section{Preliminaries}
\textbf{Long-term user interest modeling} expects to extract practical and helpful features from massive user historical behaviors for downstream CTR prediction. Let $b_i$ denote a user behavior, which refers to items that exposed or clicked by users in this paper. The $\boldsymbol{B}=[b_1,b_2,...,b_L]$ represents the historical user behavior sequence with length $L$. The long-term user interest model takes $\boldsymbol{B}$ as input and outputs a $d_h$-dimensional feature vector $\mathbf{x}_l\in\mathbb{R}^{1\times d_h}$. The $\mathbf{x}_l$, together with some other features fields, will be used in subsequent CTR prediction.

The \textbf{CTR prediction} task is typically formalized as a binary classification problem~\cite{ctr_adreview,deep_ctr}. It takes several fields of features as input, including user features, target item features, and contextual features. The CTR model outputs the probability that the target item is clicked. Formally, the training dataset is denoted as $\mathcal{D}=\{(\textbf{x}_1,y_1),...,(\textbf{x}_{|\mathcal{D}|},y_{|\mathcal{D}|})\}$, where vector $\textbf{x}_i\in \mathbb{R}^n$ represents the concatenation of multiple fields of features and $y_i\in\{0,1\}$ is the ground-truth label. $y_i=1$ indicates the user clicks the target item while $y_i=0$ means the item is not clicked. The CTR model can be formulated as a function $f:\mathbb{R}^n\rightarrow\mathbb{R}$, such that:
\begin{equation}
\hat{y} _i= f(\textbf{x}_i),
\end{equation}
where $\hat{y}_i\in[0,1]$ is the predicted click probability.

The cross-entropy loss is used to optimize the CTR model:
\begin{equation}
\mathcal{L}_{ctr} = -\frac{1}{|\mathcal{D}|}\sum_{i=1}^{|\mathcal{D}|}y_i\log \hat{y}_i+(1-y_i)\log (1-\hat{y}_i).
\end{equation}

\section{Methodology}
\subsection{Model Overview}
Figure~\ref{pic_1} illustrates the overall framework of GenLI, which consists of an interest generation module, a behavior retrieval module, and an interest fusion module. The IGM first utilizes short-term behaviors to generate multiple discrete interest distributions corresponding to different types of real-time user interests, including implicit, explicit, and relative interests. For each interest distribution, the BRM will score long-term historical behaviors via a simple lookup operation and select a subset of behaviors that best fit the corresponding type of interest. Later, the IFM separately aggregates the behaviors chosen from different interest distributions and eventually combines them to produce long-term user interest features. Next, we will introduce each part of GenLI in detail.

\subsection{Embedding Layer}
Generally, the raw input features of CTR prediction models can be categorical or numerical. For those numerical features, we transform them into categorical features by pre-processing. In this way, we could assume that all input features are in categorical form. In real industrial systems, some feature fields, such as item id, may have extremely large vocabulary sizes, reaching millions or even tens of millions. With such a huge amount, the one-hot encoding can be very sparse, leading to too many model parameters and making the training process difficult. Hence, we use an embedding layer to map categorical features to low-dimensional vectors.

Let $x$ denote one of the input categorical feature, which is an integer representing the categorical index. It is converted as:
\begin{equation}
\boldsymbol{e} = \boldsymbol{E}_{x,:},
\end{equation}
where $\boldsymbol{E}\in\mathbb{R}^{C\times d_e}$ is the embedding dictionary and each row in $\boldsymbol{E}$ represents a feature vector. The $d_e$ denotes embedding dimension and $C$ denotes the vocabulary size. $\boldsymbol{E}_{x,:}$ is the $x$-th row of $\boldsymbol{E}$. The vector $\boldsymbol{e}\in\mathbb{R}^{1\times d_e}$ is the transformed low-dimensional dense embedding.

For a user behavior $b_i$, the embedding layer transforms it to a $d$-dimensional embedding vector $\boldsymbol{e}_i\in\mathbb{R}^{d}$.

\subsection{Interest Generation Module (IGM)}
Long-term user interests are helpful in CTR prediction tasks since they reveal user preferences and habits. Constrained by limited online computing resources, extracting long-term user interests from massive historical behaviors is challenging. Most approaches use a GSU module to retrieve relevant behaviors toward the target item and discard others, reducing sequence length and eliminating noisy behaviors. Despite its effectiveness, the current target-centered GSU cannot extract full and diversified user interests because it only captures target-related interests. Meanwhile, the matching-based retrieval process increases computational burden as behavior sequence extends and ignores the behavioral interaction in sequences.

Instead of following the existing GSU workflow, we propose a novel generative-based interest generation module to produce real-time user interest distributions. This module works in a target-independent manner, allowing for the modeling of more comprehensive and diverse user interests. Based on the generated distributions, historical behaviors can obtain their scores with a lookup operation in $O(1)$ time complexity, making the process more efficient. Additionally, this generative approach can integrate behavior interaction information from sequences, further enhancing the quality of interest modeling.


Given $\boldsymbol{B}=[b_1,b_2,...,b_L]$, which denotes the user behavior sequence with length $L$. The interest generation module relies on the latest $l$ behaviors $b_1,b_2,...,b_l$, which we call short-term behaviors, to generate interest distributions that reflect the user's current real-time interests. Using short-term behaviors in the generation process enables absorbing short-term information in long-term interest modeling, fusing short-term and long-term patterns. In our design, multi-head attention (MHA)~\cite{mha,dhan} is applied to generate embedding of hidden interest. Short-term behavioral embeddings are stacked in chronological order to act as key and value in MHA, i.e. $\boldsymbol{K}_{G} =\boldsymbol{V}_{G} =[\boldsymbol{e}_{1};\boldsymbol{e}_{2};...;\boldsymbol{e}_{l}] $. Two types of query vector are used to capture various knowledge. On the one hand, we use the embedding of latest user behavior $\boldsymbol{e}_{1}\in\mathbb{R}^{1\times d}$ as the query vector for summarizing input values. On the other hand, we also add a learnable query vector $\boldsymbol{e}_{q}\in\mathbb{R}^{1\times d}$ to automatically capture the knowledge of the behavior interaction in sequences. The two types of query vector are concatenated as $\boldsymbol{Q}_{G} = [\boldsymbol{e}_{1},\boldsymbol{e}_{q}]$. With $\boldsymbol{Q}_{G}, \boldsymbol{K}_{G},$ and $ \boldsymbol{V}_{G}$, hidden interest embedding is calculated as:
\begin{equation}
\label{eq:hidden}
\boldsymbol{h} = \textsf{MHA}(\boldsymbol{Q}_{G},\boldsymbol{K}_{G},\boldsymbol{V}_{G}),
\end{equation}
where $\boldsymbol{h}\in\mathbb{R}^{1\times d_h}$ is the hidden interest embedding and $d_{h}$ is the hidden dimension. The MHA operation is defined as:
\begin{equation}
\textsf{MHA}(\boldsymbol{Q},\boldsymbol{K},\boldsymbol{V})=\textsf{concat}(\boldsymbol{h}^1,\boldsymbol{h}^2,...,\boldsymbol{h}^H)\boldsymbol{W}^o,
\end{equation}
where $H$ is the number of attention heads and $\boldsymbol{W}^{o}\in\mathbb{R}^{Hd_h\times d_h}$ is the projection parameter. The $i$-th header $\boldsymbol{h}^{i}$ is defined as:
\begin{equation}
	\boldsymbol{h}^i = \textsf{softmax}(\frac{(\boldsymbol{Q}\boldsymbol{W}_{Q}^i)(\boldsymbol{K}\boldsymbol{W}_{K}^i)^T}{\sqrt{d_h}})(\boldsymbol{V}\boldsymbol{W}_{V}^i),
\end{equation}
where $\boldsymbol{W}_{Q}^i,\boldsymbol{W}_{K}^i, \boldsymbol{W}_{V}^i$ are parameters of projection. 

Based on the extracted hidden interest embedding, the interest generation module then generates the discrete interest distribution with multi-layer perception (MLP):
\begin{equation}
\label{eq:distribution}
	\boldsymbol{p} = \textsf{softmax}(MLP(\boldsymbol{h})),
\end{equation}
where $\boldsymbol{p}\in\mathbb{R}^{1\times N}$ is the generated $N$-dimensional discrete distribution. Each entry in $\boldsymbol{p}$ is a real number between 0 and 1 that represents the probability value. 

In real-world scenarios, the user interests are highly diverse and personalized. To build a more comprehensive view of real-time user interests, GenLI framework allows us to simultaneously generate multiple distributions to indicate different aspects of user interests. Here, we present three types of real-time user interests that are conducive to CTR prediction, including implicit, explicit, and relative interests. Note that valid interests are not limited to these categories and other types of interest can be easily added to the GenLI framework depending on the needs of specific scenes, making it more practical.

The explicit user interests directly present users' current attitudes towards specific items, which can be inferred from direct interactions that users leave on the platform, such as clicks. GenLI generates a distribution $\boldsymbol{p}^E$ with hidden interest embedding $\boldsymbol{h}^E$ to capture explicit interests. $\boldsymbol{p}^E$ takes the current actual click behavior as the label and is estimated by minimizing the negative log-likelihood on probability scores of clicked target items:
 \begin{equation}
 	\mathcal{L}_{explicit} = -\sum_{i=1}^{|\mathcal{D}|}\mathbb{I}_{y_i=1}\log s(t_i,\boldsymbol{p}^E),
 \end{equation}
 where $t_i$ is the target item in the $i$-th sample. The $\mathbb{I}_{y_i=1}$ is an indicator function. It equals 1 when $y_i=1$ and 0 otherwise. The $s(t_i,\boldsymbol{p}^E)$ is the lookup probability score of item $t_i$ from distribution $\boldsymbol{p}^E$, which can be obtained via a lookup operation, i.e. $s(t_i,\boldsymbol{p}^E) =\textsf{LOOKUP}(t_i, \boldsymbol{p}^E)$. We will describe this operation in Section~\ref{sec44}.

Additionally, exposed items, which are chosen by platforms according to history logs before being displayed to users, can be viewed as a rough inference on user preferences and reflect one aspect of user interests as well. As exposed items have no direct user feedback, they are considered as implicit interests. Unlike explicit user interests, implicit interests are more stable because they are summarized in the long run. The explicit and implicit interests, respectively, emphasize user preferences at short and long periods, complementing each other in interest modeling. Hence, we generate a distribution $\boldsymbol{p}^I$ with hidden embedding $\boldsymbol{h}^I$ and minimize the negative log-likelihood on probability scores of exposed items to capture the implicit interests:
 \begin{equation}
 	\mathcal{L}_{implicit} = -\sum_{i=1}^{|\mathcal{D}|}\log s(t_i,\boldsymbol{p}^I).
 \end{equation}
 
Otherwise, we also consider the relative interest distribution, which is defined as:
\begin{equation}
\label{eq:relative}
    \boldsymbol{p}^R=\textsf{softmax}(\boldsymbol{p}^E-\boldsymbol{p}^I).
\end{equation}

Relative interest is important because it not only reflects the changing trend of user interests, but also models the user decision-making process on the platform. For one thing, explicit and implicit interests have distinct timeliness. The difference between them represents an alteration of interests, powerfully indicating the real-time user interests. For another, the decision process of a user is to find a few items they are more interested in from a pool of exposed items. Items not being clicked means that the level of interest is not high enough in comparison to other items instead of lack of interest. The relative interest can describe this contrast relation among items, thus being essential.

\subsection{Behavior Retrieval Module (BRM)}\label{sec44}
The BRM is designed to select behaviors that best fit the corresponding user interest. Given an interest distribution $\boldsymbol{p}$, a historical user behavior $b_i$ can directly obtain its score via a lookup operation:
\begin{equation}
	s(b_i,\boldsymbol{p}) =\textsf{LOOKUP}(b_i, \boldsymbol{p})=\boldsymbol{p}_{\texttt{ID}(b_i)\% N},
\end{equation}
where $\%$ is the modular operator. The function $\texttt{ID}(\cdot)$ gets the unique id of a behavior, which is prepossessed as an input feature. The overall $\texttt{ID}(b_i)\% N$ procedure transforms a behavior to the corresponding entry index within the distribution.

One historical behavior will get the score from each interest distribution, which reflects its importance under different types of user interests. To model the diverse user interests, we select behaviors with the top-$k$ score within $\boldsymbol{p}^I$, $\boldsymbol{p}^E$, and $\boldsymbol{p}^R$, respectively. We denote the corresponding selected behaviors as $\boldsymbol{B}_I$, $\boldsymbol{B}_E$ and $\boldsymbol{B}_R$. We use $K$ to denote the total retrieved behaviors, i.e., $K=|\boldsymbol{B}_I|+|\boldsymbol{B}_E|+|\boldsymbol{B}_R|=3k$.

Generally, a longer sequence of user behavior provides more information, but increases the computing burden of online services at the same time. In fact, the behavior sequence contains a lot of noise and only a few behaviors are essential for the current user interests. With BRM, GenLI could focus on the behaviors that users are most likely to be interested in and reduce the noise.

\subsection{Interest Fusion Module (IFM)}
The interest fusion module aims to facilitate interactions between retrieved behaviors and the target item, ultimately generating long-term interest features. It operates hierarchically by first separately aggregating the selected behaviors from various user interest distributions into interest embeddings. These embeddings are then combined to produce comprehensive long-term user interest features.

In detail, IFM aggregates each type of selected behaviors via multi-head attention. Without loss of generality, we take the aggregation process of $\boldsymbol{B}_I$ as an example. Let $b_1^{\prime},b_2^{\prime},...,b_k^{\prime}$ represent the selected behaviors in $\boldsymbol{B}_I$, which are sorted in temporal order. The embeddings of selected behaviors are stacked to form the key and value in MHA, that is, $\boldsymbol{K}_{I} =\boldsymbol{V}_{I} =[\boldsymbol{e}_{1}^{\prime};\boldsymbol{e}_{2}^{\prime};...;\boldsymbol{e}_{k}^{\prime}] $. The embedding of target item acts as query, i.e. $\boldsymbol{Q}_{I} =\boldsymbol{e}_{t}$. Thus, interest embedding $\boldsymbol{z}^{I} \in\mathbb{R}^{d_h}$ is calculated as:
\begin{equation}
\boldsymbol{z}^{I} = \textsf{MHA}_{F}(\boldsymbol{Q}_{I},\boldsymbol{K}_{I},\boldsymbol{V}_{I}).
\end{equation}

Similarly, the behaviors from $\boldsymbol{B}_E$ and $\boldsymbol{B}_R$ are also aggregated, and we denote the corresponding interest embedding as $\boldsymbol{z}^{E}$ and $\boldsymbol{z}^{R}$. Eventually, we concatenate three interest embeddings as $\boldsymbol{z}^F=\textsf{concat}(\boldsymbol{z}^{I},\boldsymbol{z}^{E},\boldsymbol{z}^{R})$ and produce long-term user interest features $\mathbf{x}_l$ via the gating mechanism:
\begin{equation}
\label{eq:fusing}
	\boldsymbol{g} =\sigma(MLP(\boldsymbol{z}^{F})),\quad\mathbf{x}_l = (\boldsymbol{g}\otimes\boldsymbol{z}^{F})\boldsymbol{W}^g,
\end{equation}
where $\sigma(\cdot)$ is the sigmoid activation function and $\otimes$ is the element-wise product. The $\boldsymbol{g}$ is the gating score and $\boldsymbol{W}^g$ is the projection parameter.

Besides the long-term user feature $\mathbf{x}_{l}\in\mathbb{R}^{1\times d_h}$, some other features are used for CTR prediction as well, such as short-term user interest feature $\mathbf{x}_{s}$, other side information $\mathbf{x}_o$, and embedding of target item $\boldsymbol{e}_t$. These features are concatenated as a whole vector $\mathbf{x}=\textsf{concat}(\mathbf{x}_{l},\mathbf{x}_s,\mathbf{x}_o,\boldsymbol{e}_t)$, Finally, the predicted CTR is computed with an MLP:
\begin{equation}
\hat{y} = \sigma(MLP_{ctr}(\mathbf{x})).
\end{equation}

The GenLI is optimized in an end-to-end manner with the following loss function:
\begin{equation}
\label{eq:loss}
\mathcal{L} = \mathcal{L}_{ctr}+\alpha\mathcal{L}_{implicit}+\beta\mathcal{L}_{explicit},
\end{equation}
where $\alpha$ and $\beta$ are hyper parameters to control the loss weights.

The overall workflow of GenLI is shown in Algorithm~\ref{alg:genli}.
\begin{algorithm}[!ht]
    \caption{Generative Long-term User Interest Modeling (GenLI)}
    \raggedright
    \renewcommand{\algorithmicrequire}{\textbf{Input:}}
    \renewcommand{\algorithmicensure}{\textbf{Output:}}
    \begin{algorithmic}[1]
        \label{alg:genli}
        \REQUIRE User behavior sequence $\boldsymbol{B}$.
        \ENSURE Long-term user interest feature $\mathbf{x}_l$.
        \WHILE{not converge}
        \STATE Extracting short-term behaviors $b_1,b_2,...,b_l$ from $\boldsymbol{B}$.
        \STATE Compute hidden interest embedding $\boldsymbol{h}^I,\boldsymbol{h}^E$ as in Eq.\ref{eq:hidden} with short-term behaviors, respectively.
        \STATE Compute explicit interest distribution $\boldsymbol{p}^E$ and implicit interest distribution $\boldsymbol{p}^I$ as in Eq.\ref{eq:distribution}, respectively.
        \STATE Compute relative interest distribution $\boldsymbol{p}^R$ as in Eq.\ref{eq:relative}.
        \STATE Retrieve behaviors $\boldsymbol{B}^I,\boldsymbol{B}^E,\boldsymbol{B}^R$ according to $\boldsymbol{p}^I,\boldsymbol{p}^E,\boldsymbol{p}^R$, respectively.
        \STATE Compute long-term interest feature $\mathbf{x}_l$ according to Eq.\ref{eq:fusing}.
        \STATE{Compute loss $\mathcal{L}$ as in Eq.\ref{eq:loss}} and update parameters through back-propagation.
        \ENDWHILE
        \RETURN $\mathbf{x}_l$.
    \end{algorithmic}
\end{algorithm}

\subsection{Complexity Analysis} 
For the interest generation module, the distribution generation process requires $O(ld_h)$ time, where $l$ is the length of short-term behaviors and $d_h$ is hidden dimensions. In behavior retrieval module, each behavior can get the score in $O(1)$ time, thus the total time for behavioral retrieval process is $O(L)$. For the interest fusion module, the time complexity is $O(Kd_h)$. Hence, the total complexity of GenLI is $O(L+(l+K)d_h)$. In real-world scenarios, there can be thousands of historical behaviors, while the length of short-term and retrieved behaviors are usually set to small numbers such that $ld_h\ll L$ and $Kd_h\ll L$. Thus, the complexity is around $O(L)$.

In Table~\ref{tb:complex}, we analyze the time complexity of GenLI and other methods. GenLI greatly reduces the complexity for scoring a single behavior, thus becoming less time-consuming.

\begin{table}[!ht]
	\caption{Complexity Analysis. Symbol $L$ is the length of behavior sequence, $l$ is the length of short-term behaviors, $K$ is the number of total retrieved behaviors, $d_h$ is hidden dimensions, $m$ is the length of hash fingerprint, and $M$ is size of inverted index in SIM.}
	\resizebox{\linewidth}{!}{
		\begin{tabular}{lccc}
			\toprule
			\textbf{Model}            & \textbf{Metrics}    & \textbf{Score each behavior} & \textbf{Total}   \\ \midrule
			SIM &Inner product & $O(d_h)$               & $O((M+K)d_h)$                 \\
			ETA      & Hamming distance  &   $O(m)$                & $O((L+K)m)$                  \\
			SDIM          & Hash collision  & $O(m\log(d_h))$                  & $O(Lm\log(d_h))$                  \\
			TWIN         &Target attention    & $O(d_h)$                  & $O((L+K)d_h)$                  \\
			GenLI         & Distribution lookup  & $O(1)$ & $O(L+(l+K)d_h)$              \\ \bottomrule
		\end{tabular}}
	\label{tb:complex}
\end{table}

\section{Experiment}
In this section, we conduct offline and online experiments to answer the following research questions (\textbf{RQ}s):

$\bullet$ \textbf{RQ1}: How much improvement does GenLI have on CTR prediction compared to other models?

$\bullet$ \textbf{RQ2}: What are the benefits of generative methods in GenLI?

$\bullet$ \textbf{RQ3}: How do the key hyperparameters in GenLI affect its performance?

$\bullet$ \textbf{RQ4}: How effective is GenLI in online systems?


\subsection{Datasets}
To thoroughly verify GenLI’s ability to capture long-term user interest features, we conduct experiments on widely used open-source datasets and one industrial dataset. The basic statistical information of these datasets is listed in Table~\ref{tb1:dataset}.

\begin{table}[!ht]
	\caption{Statistical information of datasets.}
	\resizebox{\linewidth}{!}{
		\begin{tabular}{lcccc}
			\toprule
			Dataset      & Users   & Items    & Categories & Instances \\ \midrule
			Amazon (Book) & 75053   & 358367   & 1583       & 150016    \\
			TaoBao       & 987994 & 34196612 & 5597       & 7956431   \\
			Industrial   & 240~million        &1.1~million          & 160           &   1.1~billion        \\ \bottomrule
	\end{tabular}}
	\label{tb1:dataset}
\end{table}

\textbf{Amazon (Book)} dataset\footnote{http://jmcauley.ucsd.edu/data/amazon/}~\cite{amazon_data} contains reviews and metadata of book products from Amazon website. We use the same pre-process techniques as in previous work. All user behaviors are arranged chronologically to form the historical behavior sequence. The length of the behavior sequence is truncated or padded to 100, acting as long-term behaviors, and the recent 10 behaviors are regarded as short-term behaviors.

\textbf{Taobao} dataset\footnote{https://tianchi.aliyun.com/dataset/649}~\cite{tb_data} collects user-item interaction records from a real-world e-commerce system named TaoBao. This dataset contains multiple types of user behaviors, such as click, purchase, favorite, etc. Following the data preprocess methods in previous works, the length of behavior sequence is truncated or padded to 500, aiming at fully demonstrating the long-term user interests modeling ability. The recent 100 behaviors are used as short-term behaviors.

\textbf{Industrial} dataset comes from advertising system of Meituan APP platform, which is the largest lifestyle services platform in China. We select consecutive 60-day samples for training and the next one day for testing. The length of behavior sequence is truncated or padded to 1000, and the recent 100 behaviors are selected as the short-term sequence.

\subsection{Baselines}
We employ the following models as the baseline models to verify the performance of GenLI:

$\bullet\ $\textbf{Avg-Pooling-long} extracts long-term user interest features via simple and fast average-pooling operations in the entire long-term user behavior sequence. The extracted long-term features are concatenated with other features to compute the final CTR via an MLP. 

$\bullet\ $\textbf{DIN}~\cite{din} applies a local activation unit to calculate the target attention between the target item and each user behavior. Later, based on the attention values, historical behaviors are aggregated to produce user interest features. Due to its high computational cost, DIN extracts only user interests from short-term user behaviors.

$\bullet\ $\textbf{DIEN}~\cite{dien} considers the user interest evolving and uses a variant GRU structure to capture evolved user interests. Similarly, DIEN only extracts short-term user interests due to its high computational complexity.

$\bullet\ $\textbf{SIM-hard $\&$ SIM-soft}~\cite{sim} first introduce the two-stage strategies into long-term user interest modeling, consisting of a GSU and an ESU module. The only difference between SIM-hard and SIM-soft lies in the GSU design. SIM-hard selects behaviors with the same category as the target item, whereas SIM-soft selects behaviors based on inner product of embeddings.

$\bullet\ $\textbf{ETA}~\cite{eta} follows the two-stage framework as well. It converts user behaviors and target items to hash signatures by SimHash algorithms. The GSU selects user behaviors according to the Hamming distance between hash signatures of target and behaviors.

$\bullet\ $\textbf{SDIM}~\cite{sdim} samples multiple hash functions to generate hash signatures of the target item and user behaviors and obtains user interest by directly gathering behaviors that have hash collisions with the target item.

$\bullet\ $\textbf{TWIN}~\cite{twin} adopts identical target-behavior attention mechanisms in GSU and ESU, which maintains the consistency of similarity metrics in two stages. It further improves the attention computing process via behavior feature splitting techniques. TWIN achieves state-of-the-art performance in capturing long-term user interest features for CTR prediction tasks.


\subsection{Experiment Setup}
To precisely assess the relative merits of different methods for modeling long-term user interests, we keep all other parts of CTR prediction the same except for the long-term user interest part. For GenLI and all baseline models, the embedding layer embeds the raw input features in a vector in 8 dimensions, that is, $d=8$. The final MLP for outputting predicted CTR has 3 layers with hidden dimensions 200 and 80, respectively. 

As for the GenLI model, the parameter $N$ is set to 4096, which means that the generated interest distributions all have 4096 entries. The MLPs for generating distributions have identical structures, using 3 layers with hidden dimensions 200 and 80 and PReLU as the activation function. All multi-head attention operations in GenLI use 4 heads and 8 hidden dimensions for each head, i.e., $H=4$ and $d_h=8$. The loss weights $\alpha$ and $\beta$ are all set to 1.0. To make a fair comparison, we keep the total number of selected behaviors the same. The baselines select 60 behaviors in GSU while GenLI selects 20 behaviors in each interest distribution, i.e., $k\text{=}20\ \text{and}\ K\text{=}60$.

During training, we use Adam optimizer with a learning rate of 0.001. The batch size is set to 256. The clicked items are used as positive samples, while not clicked items are negative samples. We maintain a positive to negative sample ratio of 1:1 via random sampling. In industrial dataset, the exposed items are recorded so that the implicit interest loss can be easily computed. For Amazon and TaoBao datasets, exposed items are unknown, thus we sample one recent similar item in behavioral sequence to represent the exposed item for calculating implicit interest loss.

In offline experiments, we use AUC as the performance metric. In online experiments, we use the click-through rate (CTR) and Revenue Per Mille (RPM) as metrics. In addition, we assess the model efficiency by comparing the average inference time on a batch of data (8192 samples). Less inference time indicates higher inference efficiency.

\subsection{Overall Performance (RQ1)}
We make an overall comparison among all models and report the results in Table~\ref{tb3:overall}. It can be observed that GenLI outperforms all other baselines on two datasets. Compared to the optimal baseline, GenLI exhibits an absolute improvement of 0.61\% and 1.03\% AUC at Amazon and TaoBao datasets, respectively, highlighting its superiority at modeling long-term user interests for CTR prediction tasks. The DIN and DIEN perform relatively poorly compared to other methods because they capture only short-term user interests. Models equipped with long-term interest features have a stronger ability to reflect user habits and preferences, thus enhancing the accuracy of CTR prediction. Additionally, two-stage models all exceed average pooling methods. These results may be attributed to the poor aggregation approach. The historical behaviors contain lots of noise and irrelevant behaviors for target items. Simply pooling brings noise, degrading the accuracy. As for the two-stage models, the proposed GenLI beats them all, achieving SOTA performance. GenLI captures implicit, explicit, and relative interests through its interest generation module, offering a more comprehensive view of user interests. Also, the multi-head attention mechanism in interest generation module integrates behavior interaction information in sequences, making the scores in retrieval process more accurate and further leading to a better performance in CTR prediction.
\begin{table}[!ht]
	\caption{Overall performance of models. The bold number indicates the best results.}
	\resizebox{\linewidth}{!}{
		\begin{tabular}{lcc}
			\toprule
			\textbf{Model}                & \textbf{Amazon (mean$\pm$std)} & \textbf{TaoBao (mean$\pm$std)}  \\ \midrule
			DIN               &   0.7267($\pm$0.0023)                & 0.8893($\pm$0.0020)                    \\
			Avg-Pooling-long &   0.7353($\pm$0.0038)                & 0.8777($\pm$0.0020)                   \\
			DIEN             &    0.7409($\pm$0.0031)               &  0.9352($\pm$0.0013)                 \\
			SIM-hard     &     0.7455($\pm$0.0032)              &   0.9403($\pm$0.0005)                 \\
			SIM-soft      &     0.7487($\pm$0.0028)              &    0.9418($\pm$0.0013)                \\
			ETA              &     0.7492($\pm$0.0021)              &  0.9438($\pm$0.0005)                \\
			SDIM            &     0.7429($\pm$0.0028)              &  0.9405($\pm$0.0009)                  \\
			TWIN            &     0.7503($\pm$0.0004)              &  0.9449($\pm$0.0002)                  \\ 
                \rowcolor{gray!40}
			\textbf{GenLI (Ours)}           & \textbf{0.7564($\pm$0.0012)} & \textbf{0.9552($\pm$0.0006) }             \\ \bottomrule
		\end{tabular}}
	\label{tb3:overall}
\end{table}

In industrial dataset, we compare GenLI with the most common baselines and the current SOTA baseline. The results are listed in Table~\ref{tb4:industrial}. GenLI improves the AUC by about 0.22\% compared to the TWIN. Note that even a 0.1\% increase in offline experiments is significant enough to increase online revenue. Otherwise, GenLI has advantages in inference efficiency. GenLI costs slightly more time than average pooling but gains more AUC improvements. Compared with TWIN and SIM, GenLI gains better AUC and reduces the inference time.
\begin{table}[!ht]
	\caption{CTR performance of models on industrial dataset.}
		\begin{tabular}{lcc}
			\toprule
			\textbf{Model}                & \textbf{AUC} & \textbf{Inference Time(ms)}  \\ \midrule
			Avg-Pooling-long &   0.7424               & \textbf{2.8}                   \\
			SIM-hard     &   0.7437                & 5.6                 \\
			SIM-soft      &   0.7432                & 6.8                   \\
			TWIN            & 0.7441                  &  7.9                \\
			\rowcolor{gray!40}\textbf{GenLI(Ours)}           & \textbf{0.7463} & 4.6              \\ \bottomrule
		\end{tabular}
	\label{tb4:industrial}
\end{table}

\subsection{Effects of Interest Generation Module (RQ2)}
\begin{figure}[!ht]
	\centering
	\includegraphics[width=\linewidth]{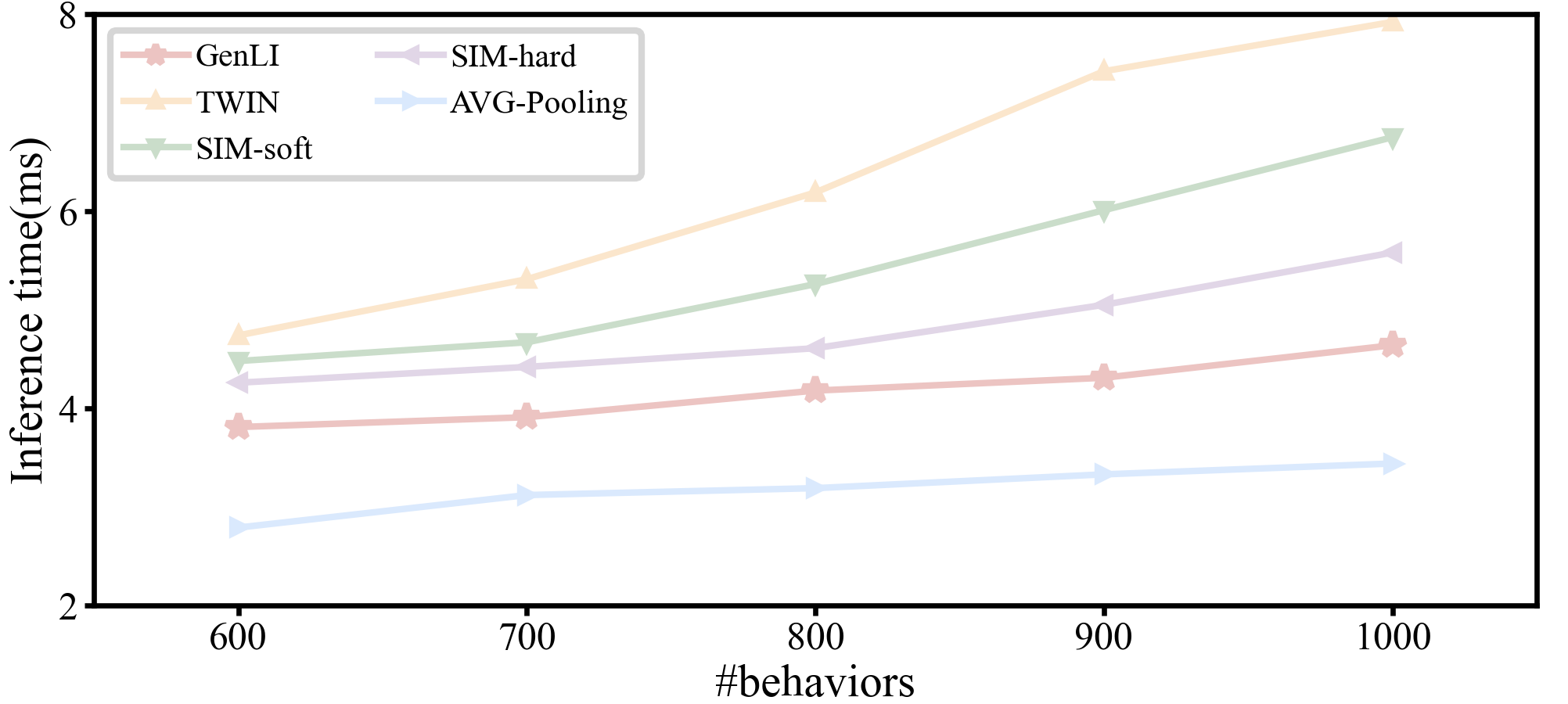}
	\caption{Inference time comparison of models.}
	\label{exp:pic_time}
\end{figure}

To fully understand the impact of generation methods on performance, we conduct experiments on core parts of the interest generation module.

Firstly, we discuss its advantages in inference efficiency. Fig~\ref{exp:pic_time} shows the inference time on different lengths of user historical behaviors. It can be noticed that the inference time of GenLI is significantly lower than that of TWIN and SIM at all levels of behavior sequence length. Meanwhile, when the length increases, efficiency improvements of GenLI are more obvious, which might be attributed to the lower complexity for scoring each behavior.


Also, we discuss the impacts of different types of generated distributions in interest generation module. We remove some distributions and list the results in Table~\ref{tb3:distribution}. The GenLI \textit{w/o} IGM model means totally remove the IGM and selecting behaviors almost randomly. GenLI \textit{w/o} IGM shows an obvious performance drop, indicating the importance of modeling interest distributions. Meanwhile, we observe that solely without $\boldsymbol{p}^I$, $\boldsymbol{p}^E$ or $\boldsymbol{p}^R$, the AUC score will all decrease. This result proves that all three generated distributions are helpful for long-term user interest modeling. Moreover, we also notice that GenLI without $\boldsymbol{p}^E$ has the worst performance. This phenomenon indicates that modeling explicit interest distribution has a more direct influence on CTR prediction performance.

\begin{table}[!ht]
	\caption{Effects of generated distributions on performance.}
		\begin{tabular}{lcc}
			\toprule
			\textbf{Model}      & \textbf{Amazon}   & \textbf{TaoBao}  \\ \midrule
            GenLI \textit{w/o} IGM & 0.7361(-2.68\%) & 0.9227(-3.40\%) \\
			GenLI \textit{w/o} $\boldsymbol{p}^I$       & 0.7465 (-1.31\%)   & 0.9413 (-1.46\%)      \\
			GenLI \textit{w/o} $\boldsymbol{p}^E$      & 0.7400 (-2.17\%)   & 0.9372 (-1.88\%)       \\
   GenLI \textit{w/o} $\boldsymbol{p}^R$      & 0.7532 (-0.42\%)   & 0.9526 (-0.27\%)       \\
			GenLI       & 0.7564 & 0.9552    \\ \bottomrule
		\end{tabular}
	\label{tb3:distribution}
\end{table}

\begin{figure}[!ht]
	\centering
	\includegraphics[width=\linewidth]{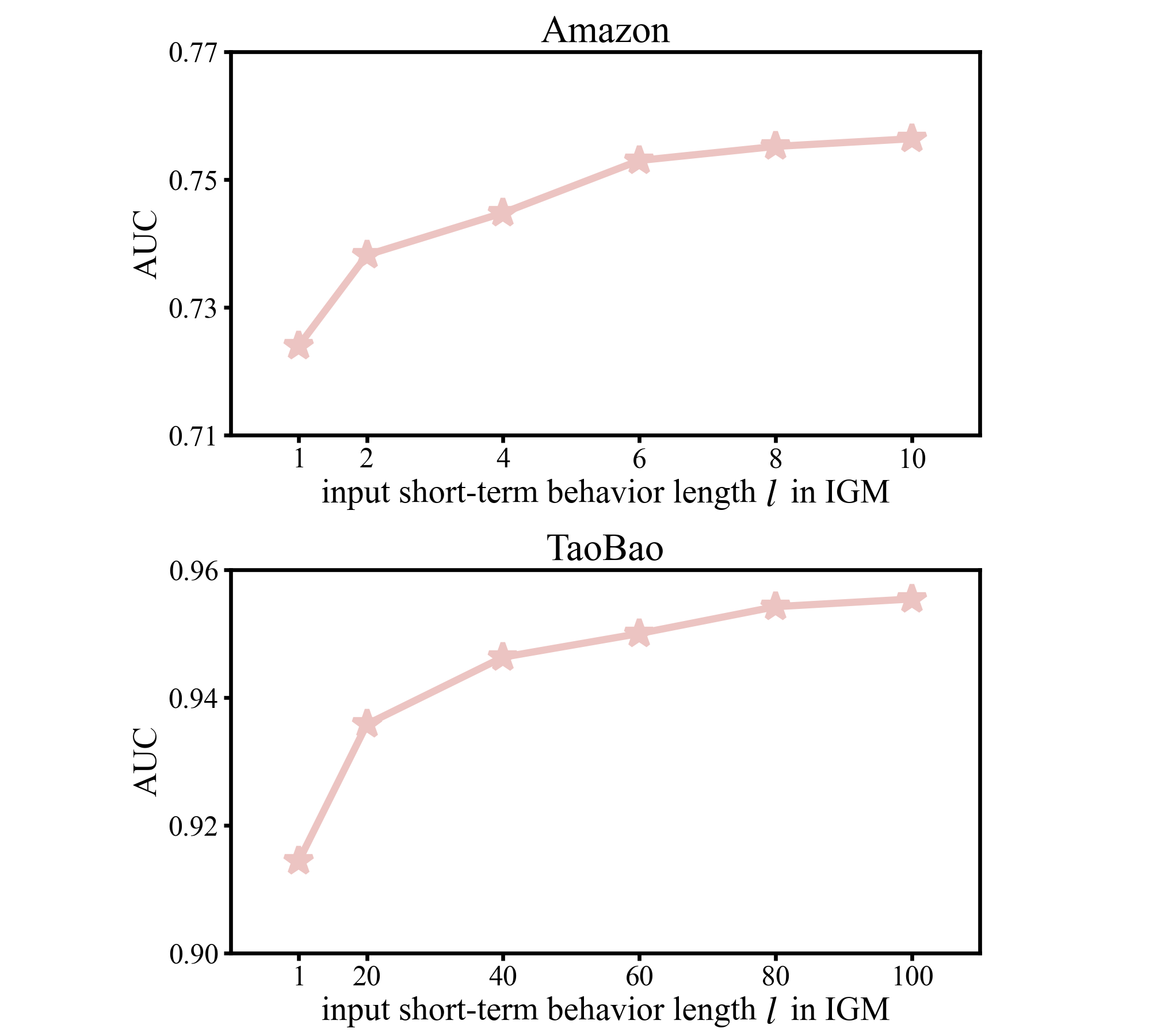}
	\caption{Performance of GenLI when taking different lengths of short-term behaviors as the input of interest generation module.}
	\label{exp:pic_2}
\end{figure}

In the IGM, short-term behaviors are used to generate interest distributions. We vary the length of input short-term behaviors and compare the performance in Fig~\ref{exp:pic_2}. For both two datasets, the AUC increases as the behavioral length grows. Longer behavior input provides more interaction patterns, enabling a better score for the retrieval process. Although increasing the length of input behaviors can enhance the modeling of current interests, the improvement becomes limited after inputs reach a certain size, whereas the computational cost increases sharply. Therefore, using short-term behaviors in IGM is a more suitable choice.

\subsection{Ablation Study (RQ3)}

We conduct experiments on key hyperparameters $K$ and $N$ in GenLI to provide an in-depth analysis of their impact on model performance.

\begin{figure}[!ht]
	\centering
	\includegraphics[width=\linewidth]{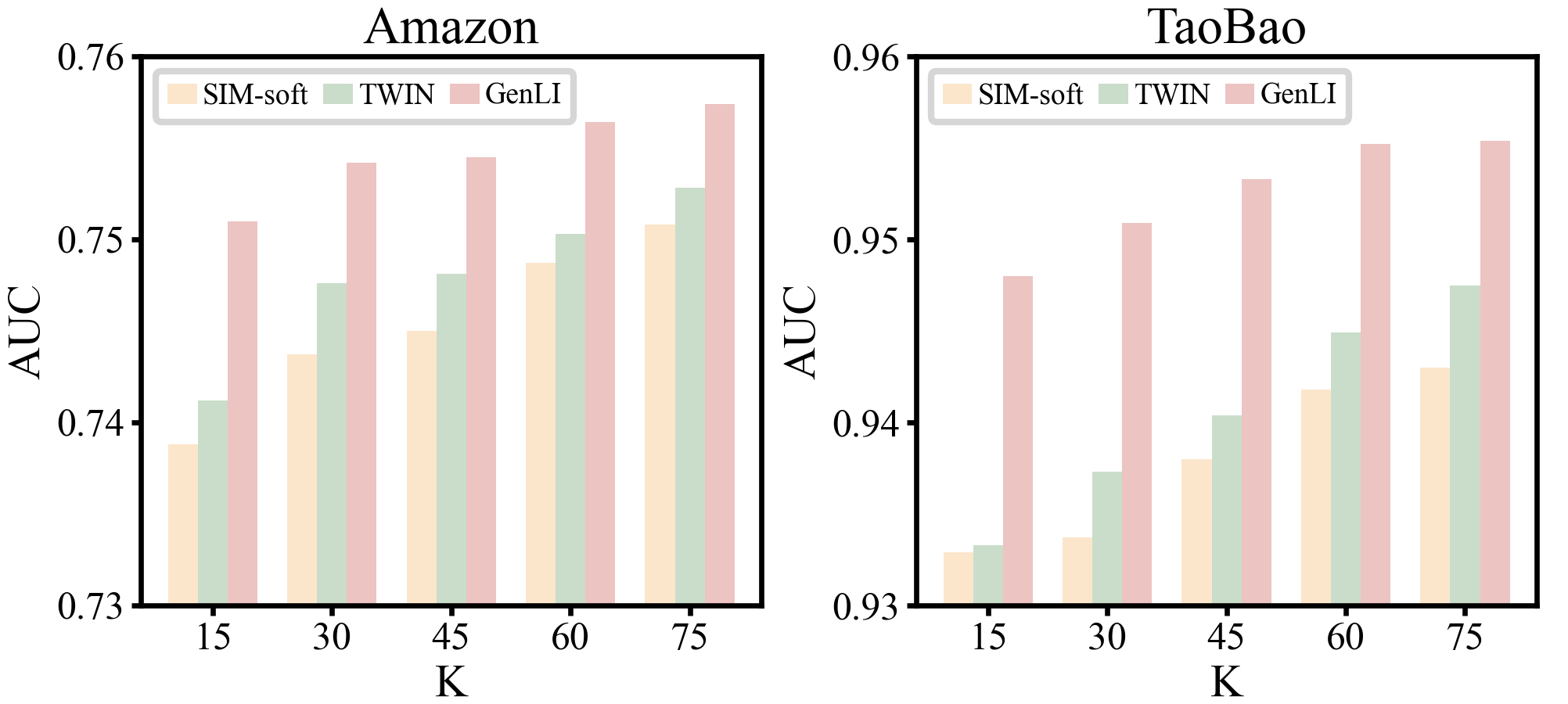}
	\caption{Performance comparison when retrieving different numbers of historical behaviors.}
	\label{exp:pic_4}
\end{figure}

\begin{figure}[!ht]
	\centering
	\includegraphics[width=\linewidth]{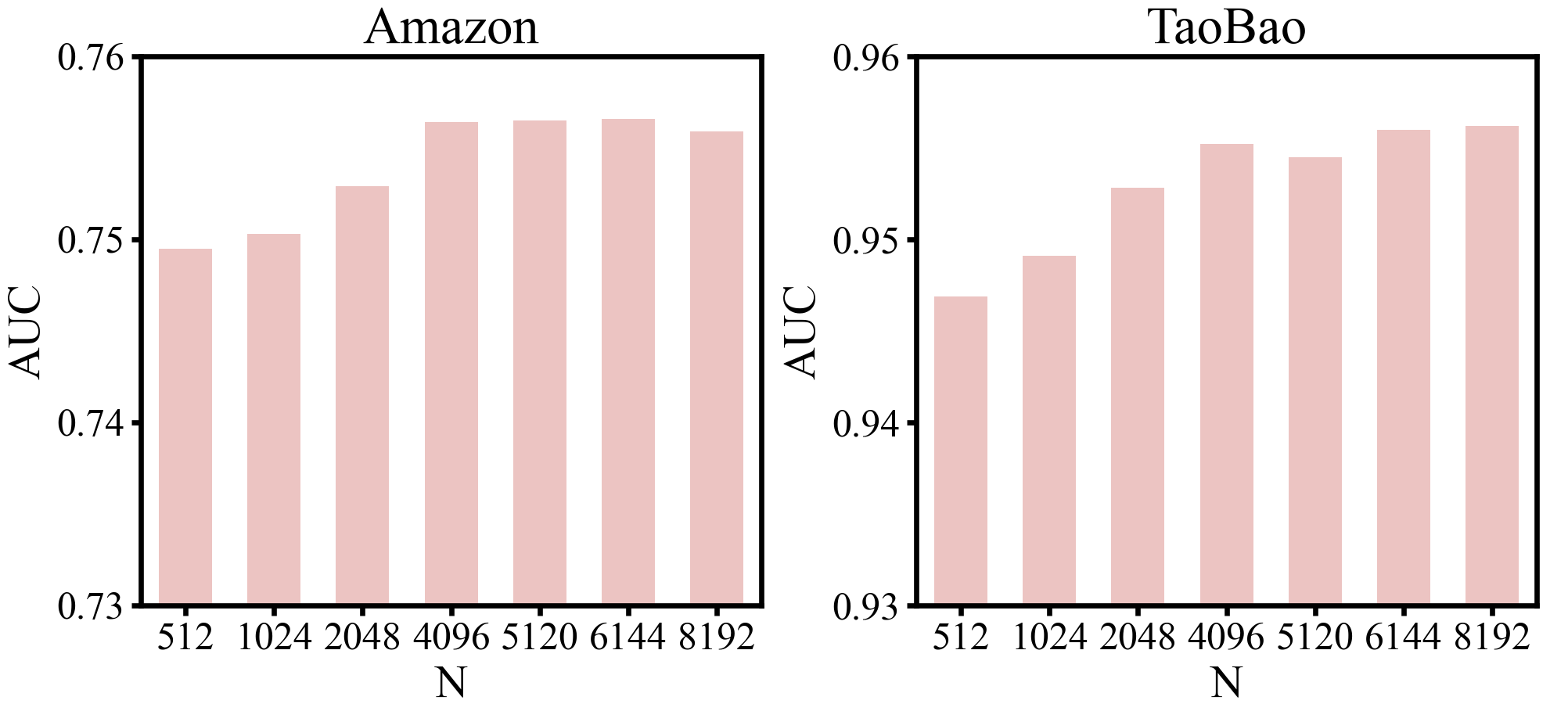}
	\caption{Performance of GenLI under different generated distribution dimensions.}
	\label{exp:pic_5}
\end{figure}

The parameter $K$ controls the total number of behaviors selected in the BRM. Generally, increasing $K$ values moderately admits more relevant behaviors in historical behavior sequence to participate in subsequent interest fusion module, bringing more diverse interest patterns and leading to a higher prediction accuracy. However, too large $K$ would take many noise behaviors and increase time consumption. We compare the performance of SIM-soft, TWIN, and GenLI under different $K$ settings. The results are displayed in Fig~\ref{exp:pic_4}. Accordingly, for all three models, AUC first increases with $K$ rises and finally reaches a plateau as $K$ creeps up to 60. GenLI continuously surpasses SIM-soft and TWIN when $K$ varies from 15 to 75. This outcome indicates that behaviors selected by GenLI are more consistent with user real-time interests, thus contributing to superior long-term user interest modeling.

Interestingly, we find that GenLI has more significant AUC improvements at lower $K$ levels than other models. When $K=15$ and compared to TWIN, GenLI increases AUC by about 1.65\% and 1.62\% in the two datasets, respectively. While $K=75$, the improvements are about 0.88\% and 1.31\%. This phenomenon occurs because increasing the number of retrieved behaviors leads to broader coverage of user interests while reducing the significance of retrieval accuracy. Supposing an extreme situation where all historical behaviors are retrieved, all methods will have the same outcome. Therefore, when the number of retrieved behaviors is limited, the significant improvement achieved by GenLI demonstrates its superiority in behavior retrieval. From another perspective, target-centered retrieval methods can only retrieve items that are very similar to the target, failing to capture the user's complete and diverse interests. The IGM is target-independent and fully leverages the behavioral interaction information in short-term behaviors, thus more comprehensively capturing users' multiple interests. This result further reveals the limitations of target-centered retrieval methods and validates the comprehensiveness of GenLI in interest modeling.

Fig~\ref{exp:pic_5} shows the variation trend of AUC with hyperparameter $N$. The parameter $N$ is the total entries of generated discrete distributions. The figure shows that the AUC score slightly increases when parameter $N$ increases and later remains stable. Note that we use modular operation to transform behaviors to corresponding entries in distributions, which is a hashing process, in essence. More entries will reduce the collision probability, making the score more reliable. As a result, expanding $N$ can improve performance. As $N$ continues to increase, the performance improvement becomes insignificant. The final choice for the suitable $N$ is 4096 to balance performance and effectiveness.

\subsection{Online A/B Testing (RQ4)}
An online A/B testing was conducted by deploying GenLI to handle real 30\% traffic in the Meituan food delivery platform for six days. The baseline model was the online-serving CTR model of the Meituan food delivery platform, which has the leading performance in similar applications. The results are shown in Table~\ref{tb3:abtest}, demonstrating that GenLI increased the CTR by 0.776\% and the RPM by 1.567\% in comparison to the base model. Currently, GenLI has been deployed in the advertisement system of a real-world platform and is serving hundreds of millions of users, contributing to a notable increase in business revenue.

\begin{table}[!ht]
	\caption{Online A/B testing performance.}
		\begin{tabular}{lcc}
			\toprule
			Model      & CTR Gain   & RPM Gain  \\ \midrule
			Base & 0   & 0       \\
			GenLI       & +0.776\% & +1.567\%    \\ \bottomrule
		\end{tabular}
	\label{tb3:abtest}
\end{table}

\section{Conclusion}
In this paper, we propose a long-term user interest model named GenLI for CTR prediction. GenLI uses a target-independent interest generation module to generate complete and diverse user interest distributions. For each interest distribution, the behavior retrieval module will score long-term historical behaviors via a simple lookup operation and select a subset of behaviors that best fit the corresponding type of interest. Later, the interest fusion module separately aggregates the selected behaviors and eventually produces the long-term user interest features. During offline experiments, GenLI gains notable improvements. In online A/B testing, GenLI achieves 0.776\% CTR and 1.567\% RPM promotion. Currently, GenLI is deployed in online systems and serves millions of users every day.

\bibliographystyle{ACM-Reference-Format}
\bibliography{sample-base}


\end{document}